\documentstyle[prl,aps,epsfig,multicol]{revtex}
\newcommand{\beq}{\begin{equation}}
\newcommand{\eeq}{\end{equation}}

\begin{document}

\title{Nonlinear, ground-state, pump-probe spectroscopy}
\author{P. R. Berman and B. Dubetsky}

\address{Physics Department, University of Michigan, Ann Arbor, MI 48109-1120}

\date{\today}
\maketitle

\begin{abstract}
A theory of pump-probe spectroscopy is developed in which optical fields
drive two-quantum, Raman-like transitions between ground state sublevels.
Three fields are incident on an ensemble of atoms. Two of the fields act as
the pump field for the two-quantum transitions. The absorption or gain of an
additional probe field is monitored as a function of its detuning from one
of the fields which constitutes the pump field. Although the probe
absorption spectrum displays features common to those found in pump-probe
spectroscopy of single-quantum transitions, new interference effects are
found to modify the spectrum. Many of these features can be explained within
the context of a dressed atom picture.
\end{abstract}

\pacs{32.80.-t, 42.65.-k, 32.70.Jz}

\begin{multicols}{2}

\section{Introduction}

Of fundamental interest in nonlinear spectroscopy is the response of an
atomic vapor to the simultaneous application of a pump and a probe field. A
calculation of the probe field absorption is relatively straightforward \cite
{mol,har} in the weak probe field limit. Let $\Omega $ and $\Omega ^{\prime
}\ $denote the pump and probe field frequencies, $\Delta =\Omega -\omega $
the pump field detuning from atomic resonance $\omega $, and $\delta
_{1}=\Omega ^{\prime }-\Omega $ the probe-pump detuning. For a pump field
detuning $\left| \Delta \right| \gg \gamma _{e}$, $\chi $, where $\gamma
_{e} $ is the upper state decay rate and $\chi $ is a pump-field Rabi
frequency, one finds the spectrum to consist of three components. There is
an absorption peak centered near $\delta _{1}=-\Delta $ ($\Omega ^{\prime
}=\omega ),$ an emission peak centered near $\delta _{1}=\Delta $ ($\Omega
^{\prime }=2\Omega -\omega )$ and a dispersive like structure centered near $%
\delta _{1}=0$. Experimentally, a spectrum exhibiting all these features was
first obtained by Wu {\it et al.} \cite{wu}. The absorption and emission
peaks can be given a simple interpretation in a dressed-atom picture \cite
{cohen}, but the non-secular structure centered at $\delta _{1}=0$ is
somewhat more difficult to interpret \cite{gryn,berm}. The width of these
spectral components is on the order of $\gamma _{e}$, neglecting any Doppler
broadening.

The spectral response can change dramatically when atomic recoil
accompanying the absorption or emission of radiation becomes a factor \cite
{recoil}, as in the case of a highly collimated atomic beam or for atoms
cooled below the recoil limit. In this limit, the absorption and emission
peaks are each replaced by an absorption-emission doublet, and the
dispersive-like structure is replaced by a pair of absorption-emission
doublets. The spectrum can be given a simple interpretation in terms of a
dressed atom theory, including quantization of the atoms' center-of-mass
motion \cite{recoil}. It turns out, however, that at most one
absorption-emission doublet (one of the central ones) can be resolved unless
the excited state decay rate is smaller than the recoil shift. Since this
condition is violated for allowed electronic transitions, it is of some
interest to look for alternative level schemes in which this structure can
be resolved fully. If the optical transitions are replaced by two-photon,
Raman-like transitions between ground state levels, the widths of the
various spectral components are determined by ground state relaxation rates,
rather than excited state decay rates. As a result, the probe's spectral
response should be fully resolvable. Raman processes have taken on added
importance in sub-Doppler \cite{3} and sub-recoil \cite{4} cooling, atom
focusing \cite{focus}, atom interferometry \cite{5,6,7,8}, and as a method
for probing Bose condensates \cite{bose}.

In this article we propose a scheme for pump-probe spectroscopy of an atomic
vapor using Raman transitions. This is but one of a class of interactions
that can be considered under the general heading of {\em nonlinear ground
state spectroscopy}. The spectral response is found to be similar to that of
traditional pump-probe spectroscopy \cite{mol}; however, new interference
phenomena can modify the spectrum [Sec. III]. The interference phenomena can
be interpreted in terms of a dressed atom picture [Sec. IV]. Although part
of the motivation for this work is the study of recoil effects, such effects
are neglected in this article.

\section{Equations of Motion}

\begin{figure}
\begin{minipage}{.5\linewidth}
\begin{center}
\epsfxsize=\linewidth \epsfbox{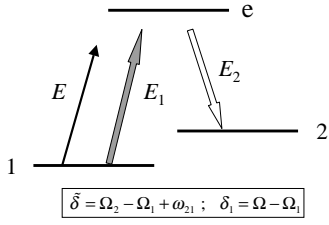}
\end{center}
\end{minipage}
\begin{minipage}{.4\linewidth} \caption{Schematic diagram 
of the atom-field 
system. Fields $E_{1}$ and $E$
drive only the $1-e$ transition and field $E_{2}$ only the $2-e$ transition. 
\label{fig1}}
\end{minipage}
\end{figure}

The atom field geometry is indicated schematically in Fig. \ref{fig1}. 
Three-level
atoms interact with two optical fields, $E_{1}$ and $E_{2}$, producing
strong coupling between initial and final levels $1$ and $2$ via an
intermediate excited state level $e$. 
Field $E_{1}$ couples only levels $1$
and $\ e$, while field $E_{2}$ couples only levels $2$ and $\ e$. In
addition, there is a weak probe field $E$ that couples only levels $1$ and $%
\ e$. As a consequence, fields $E$ and $E_{2}$ can also drive two-photon
transitions between levels $1$ and $2$. 
Levels $1$ and $2$ are pumped
incoherently at rates $\Lambda _{1}$ and $\Lambda _{2}$, respectively, and
both states decay at rate $\Gamma .$ The incoherent pumping and decay
represent an oversimplified model for atoms entering and leaving the
interaction volume.
The incident fields are assumed to be nearly
copropagating so that all two-photon Doppler shifts can be neglected. In
this limit and in the limit of large detuning on each single photon
transition, one can consider the atoms to be stationary with regards to
their interaction with the external fields. We wish to calculate the linear
probe absorption spectrum.
The electric field can be written as

\end{multicols}

\begin{equation}
E\left( {\bf R,}t\right) =\frac{1}{2}\left[ E_{1}e^{i\left( {\bf k}_{1}\cdot 
{\bf R}-\Omega _{1}t\right) }+E_{2}e^{i\left( {\bf k}_{2}\cdot {\bf R}%
-\Omega _{2}t\right) }\right. +\left. Ee^{i\left( {\bf k}\cdot {\bf R}%
-\Omega t\right) }\right] +c.c.,  \label{111}
\end{equation}

\begin{multicols}{2}

where $\Omega _{1},$ $\Omega _{2}$, and $\Omega $ are the field frequencies, 
${\bf k}_{1},$ ${\bf k}_{2}$, and ${\bf k}$ the field propagation vectors,
and $c.c.$ stands for complex conjugate. In an interaction representation,
neglecting any decay or incoherent pumping of the ground state levels, the
state probability amplitudes obey the equations of motion. 
\begin{mathletters}
\label{2}
\begin{eqnarray}
i\dot{a}_{e} &=&\chi _{1}e^{-i\Delta _{1}t}a_{1}+\chi _{2}e^{-i\Delta
_{2}t}a_{2}+\chi e^{-i\Delta t}a_{1}  \nonumber \\
&&-i\left( \gamma _{e}/2\right) a_{e},  \label{2a} \\
i\dot{a}_{1} &=&\chi _{1}e^{i\Delta _{1}t}a_{e}+\chi e^{i\Delta t}a_{e},
\label{2b} \\
i\dot{a}_{2} &=&\chi _{2}e^{i\Delta _{2}t}a_{e},  \label{2c}
\end{eqnarray}
where $\chi _{j}=-d_{ej}E_{j}/2\hbar $ ($j=1,2)$ and $\chi =-d_{e1}E/2\hbar $
are Rabi frequencies (assumed to be real and positive), $d_{ej}$ is a dipole
moment matrix element, and $\Delta _{j}=\Omega _{j}-\omega _{ej}$ and $%
\Delta =\Omega -\omega _{e1}$ are atom-field detunings$.$ Assuming that the
magnitude of the detunings are much larger than $\gamma _{e}$ and any
Doppler shifts associated with the single photon transitions, it is possible
to adiabatically eliminate the excited state amplitude to arrive at the
following equations for the ground state amplitudes: 
\end{mathletters}
\begin{mathletters}
\label{3}
\begin{eqnarray}
i\dot{a}_{1} &=&S_{1}a_{1}+S\left( e^{i\delta _{1}t}+e^{-i\delta
_{1}t}\right) a_{1}+ge^{-i\tilde{\delta}t}a_{2}  \nonumber \\
&&+g^{\prime }e^{-i\delta ^{\prime }t}a_{2};  \label{3a} \\
i\dot{a}_{2} &=&S_{2}a_{2}+ge^{i\tilde{\delta}t}a_{1}+g^{\prime }e^{i\delta
^{\prime }t}a_{1},  \label{3b}
\end{eqnarray}
where 
\end{mathletters}
\begin{mathletters}
\label{4}
\begin{eqnarray}
\tilde{\delta} &=&\Delta _{2}-\Delta _{1}=\Omega _{2}-\Omega _{1}+\omega
_{21};  \label{4a} \\
\delta ^{\prime } &=&\Delta _{2}-\Delta =\Omega _{2}-\Omega +\omega _{21};
\label{4b} \\
\delta _{1} &=&\Delta -\Delta _{1}=\Omega -\Omega _{1}=\tilde{\delta}-\delta
^{\prime },  \label{4c}
\end{eqnarray}
are detunings associated with two-quantum processes and 
\end{mathletters}
\begin{eqnarray}
g &=&\chi _{1}\chi _{2}/\Delta ;\text{ \ }g^{\prime }=\chi \chi _{2}/\Delta ;%
\text{ \ }  \nonumber \\
S_{1} &=&\chi _{1}^{2}/\Delta ;\text{ \ }S_{2}=\chi _{2}^{2}/\Delta ;\text{
\ }S=\chi \chi _{1}/\Delta ,  \label{5}
\end{eqnarray}
are Rabi frequencies or Stark shifts associated with two quantum processes.
In writing Eqs. (\ref{3}), we assumed that $\Delta \approx \Delta
_{1}\approx \Delta _{2}$ and $\left| \Delta \right| \gg \left| \tilde{\delta}%
\right| ,\left| \delta ^{\prime }\right| ,\left| \delta _{1}\right| .$

It will prove convenient, especially when going over to a dressed atom
picture, to introduce a representation in which 
\begin{eqnarray}
a_{1} &=&b_{1}e^{-i\tilde{\delta}t/2}e^{-i(S_{1}+S_{2})t/2};  \nonumber \\
a_{2} &=&b_{2}e^{i\tilde{\delta}t/2}e^{-i(S_{1}+S_{2})t/2}.  \label{6}
\end{eqnarray}
Combining Eqs. (\ref{3}) and (\ref{6}) one finds 
\begin{mathletters}
\label{7}
\begin{eqnarray}
i\dot{b}_{1} &=&-(\delta /2)b_{1}+gb_{2}+S\left( e^{i\delta
_{1}t}+e^{-i\delta _{1}t}\right) b_{1}  \nonumber \\
&&+g^{\prime }e^{i\delta _{1}t}b_{2};  \label{7a} \\
i\dot{b}_{2} &=&(\delta /2)b_{2}+gb_{1}+g^{\prime }e^{-i\delta _{1}t}b_{1},
\label{7b}
\end{eqnarray}
where 
\end{mathletters}
\begin{equation}
\delta =\tilde{\delta}-(S_{1}-S_{2}).  \label{8}
\end{equation}
The corresponding equations for density matrix elements $\rho _{11}=\left|
b_{1}\right| ^{2}$, $\rho _{22}=\left| b_{2}\right| ^{2}$, $\rho
_{12}=b_{1}b_{2}^{\ast }=\rho _{21}^{\ast }$ are

\end{multicols}

\begin{mathletters}
\label{9}
\begin{eqnarray}
\dot{\rho}_{11} &=&-ig\left( \rho _{21}-\rho _{12}\right) -ig^{\prime
}e^{i\delta _{1}t}\rho _{21}+ig^{\prime }e^{-i\delta _{1}t}\rho _{12}-\Gamma
\rho _{11}+\Lambda _{1};  \label{9a} \\
\dot{\rho}_{22} &=&ig\left( \rho _{21}-\rho _{12}\right) +ig^{\prime
}e^{i\delta _{1}t}\rho _{21}-ig^{\prime }e^{-i\delta _{1}t}\rho _{12}-\Gamma
\rho _{22}+\Lambda _{2};  \label{9b} \\
\dot{\rho}_{12} &=&i\delta \rho _{12}-ig\left( \rho _{22}-\rho _{11}\right)
-ig^{\prime }e^{i\delta _{1}t}\left( \rho _{22}-\rho _{11}\right) -iS\left(
e^{i\delta _{1}t}+e^{-i\delta _{1}t}\right) \rho _{12}-\Gamma \rho _{12},
\label{9c}
\end{eqnarray}
\end{mathletters}

\begin{multicols}{2}

where the incoherent pumping and decay terms have been introduced. It is
important to note that, in this representation, the frequency appearing in
the $g^{\prime }$ terms is $\delta _{1}=\delta ^{\prime }-\tilde{\delta}%
=\Omega -\Omega _{1}$. In other words, the effective field frequency
associated with field $E_{2}$ in this representation is $\Omega _{1}$ rather
than $\Omega _{2}$.

It follows from the Maxwell-Bloch equations that the probe absorption
coefficient, $\alpha $, and index change, $\Delta n$, are given by 
\begin{mathletters}
\label{10}
\begin{eqnarray}
\alpha &=&\frac{kNd_{1e}^{2}}{2\hbar \epsilon _{0}}%
\mathop{\rm Im}%
\left( \frac{\rho _{1e}^{\prime }}{\chi }\right) ;  \label{10a} \\
\Delta n &=&-\frac{Nd_{1e}^{2}}{2\hbar \epsilon _{0}}%
\mathop{\rm Re}%
\left( \frac{\rho _{1e}^{\prime }}{\chi }\right) ,  \label{10b}
\end{eqnarray}
where $N$ is the atomic density, 
\end{mathletters}
\begin{equation}
\rho _{1e}^{\prime }\approx \frac{1}{\Delta }\left[ \chi \rho
_{11}^{(0)}+\chi _{1}\rho _{11}^{+}+\chi _{2}\rho _{12}^{+}\right] ,
\label{11}
\end{equation}
and $\rho _{11}^{(0)}$, $\rho _{11}^{+}$, and $\rho _{12}^{+}$ are
coefficients that appear in the solution of Eqs. (\ref{9}) (to first order
in $\chi $) written in the form: 
\begin{equation}
\rho _{jj^{\prime }}=\rho _{jj^{\prime }}^{(0)}+\rho _{jj^{\prime
}}^{+}e^{i\delta _{1}t}+\rho _{jj^{\prime }}^{-}e^{-i\delta _{1}t};\text{ \
\ \ }j,j^{\prime }=1,2  \label{12}
\end{equation}
The first and third terms in Eq. (\ref{11}) are analogous to the terms that
appear in conventional theories of pump-probe spectroscopy, but the second
term is new and leads to qualitatively new features in the probe absorption
spectrum.

An expression for $\rho _{1e}^{\prime }$ is given in Appendix A. The
absorption coefficient is plotted in Figs. \ref{fig2}(a)-(c) for several values of $%
\delta /g$, and 
\begin{equation}
\eta =\sqrt{\chi _{1}/\chi _{2}}.  \label{13}
\end{equation}
If $\eta \ll 1$, the two-quantum probe absorption spectrum has the same
structure as the probe absorption spectrum involving single quantum
transitions. The situation changes if $\eta \gtrsim 1$. For example, aside
from an interchange of absorption and gain components as a function of $%
\delta $, the probe spectrum for single quantum transitions depends only on
the magnitude of the pump field detuning. This is clearly {\em not} the case
for two-quantum transitions, as is evident from Fig. \ref{fig2}(a) drawn for $\eta
=1, $ $\Gamma /g=0.1$, $\delta /g=\pm 1$. Probe absorption and gain {\em are}
interchanged when $\delta $ changes sign, but the ratio of the amplitude of
the absorption to gain peak {\em changes }when $\delta $ changes sign. There
is another subtle difference present in these spectra. The sense of the
central dispersive component is opposite to that for single quantum
transitions. With decreasing $\eta $ , the sense of the central component
would reverse, as the spectrum reverts to the same structure found in
pump-probe spectroscopy of single quantum transitions. The probe response
also depends on the sign of $\Delta $ (through $g=\chi _{1}\chi _{2}/\Delta $%
); this feature follows from the dependence of the spectrum on the sign of $%
\delta $ and the relationship 
\begin{equation}
\rho _{1e}^{\prime }(-\delta ,-\Delta ,-\delta _{1})=-\rho _{1e}^{\prime
}(\delta ,\Delta ,\delta _{1})^{\ast },  \label{reflect}
\end{equation}
which can be derived using Eqs. (\ref{A3})-(\ref{A7}) of Appendix A. It is
also possible for the components centered at positive or negative $\delta
_{1}$ to vanish (in the secular approximation) for certain values of $\eta $%
, as can be seen in Fig. \ref{fig2}(b).
\begin{figure}
\begin{minipage}{.95\linewidth}
\begin{center}
\epsfxsize=.95\linewidth \epsfysize=.68\linewidth \epsfbox{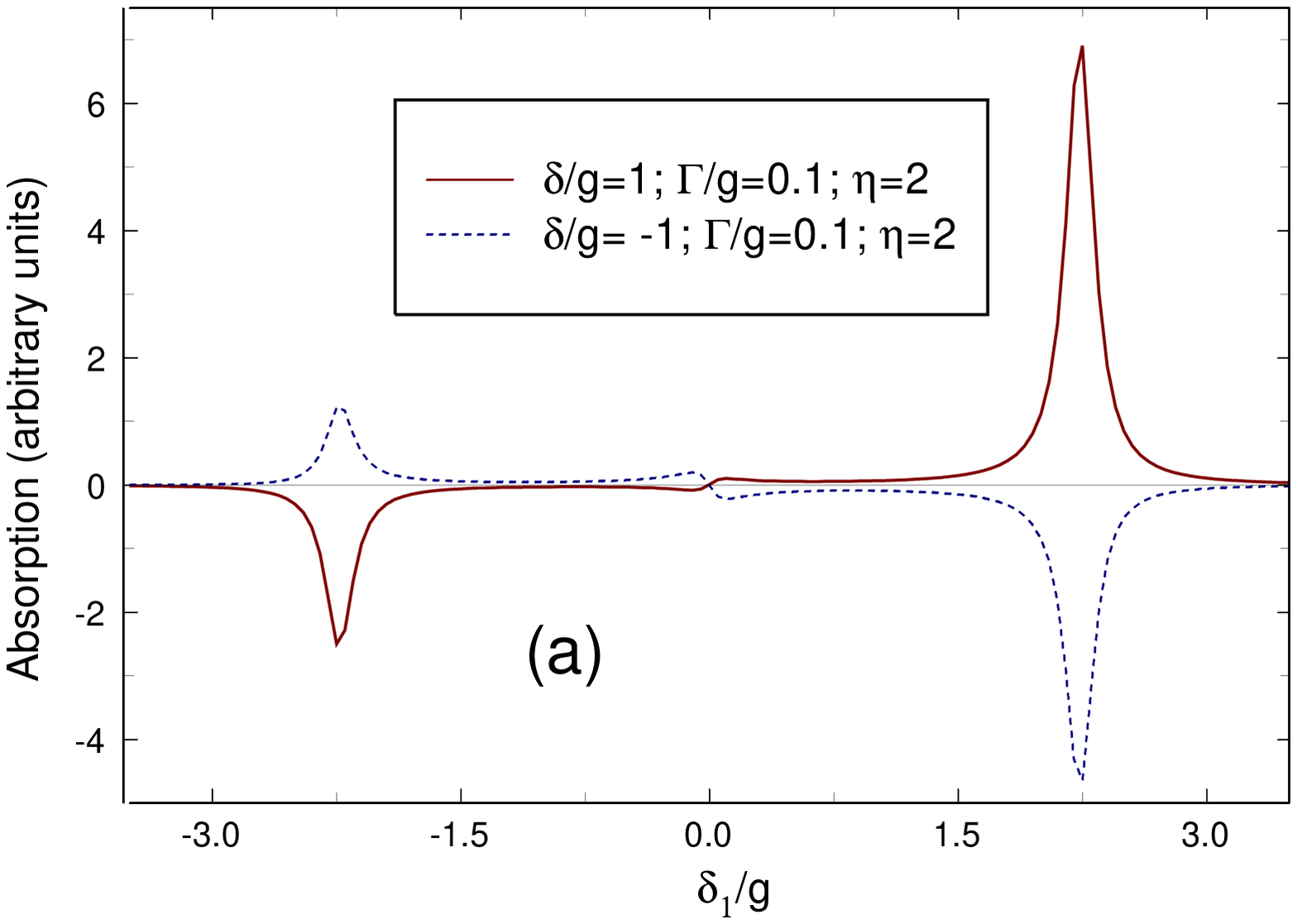}
\end{center}
\end{minipage}
\end{figure}
\begin{figure}
\begin{minipage}{.95\linewidth}
\begin{center}
\epsfxsize=.95\linewidth \epsfysize=.68\linewidth \epsfbox{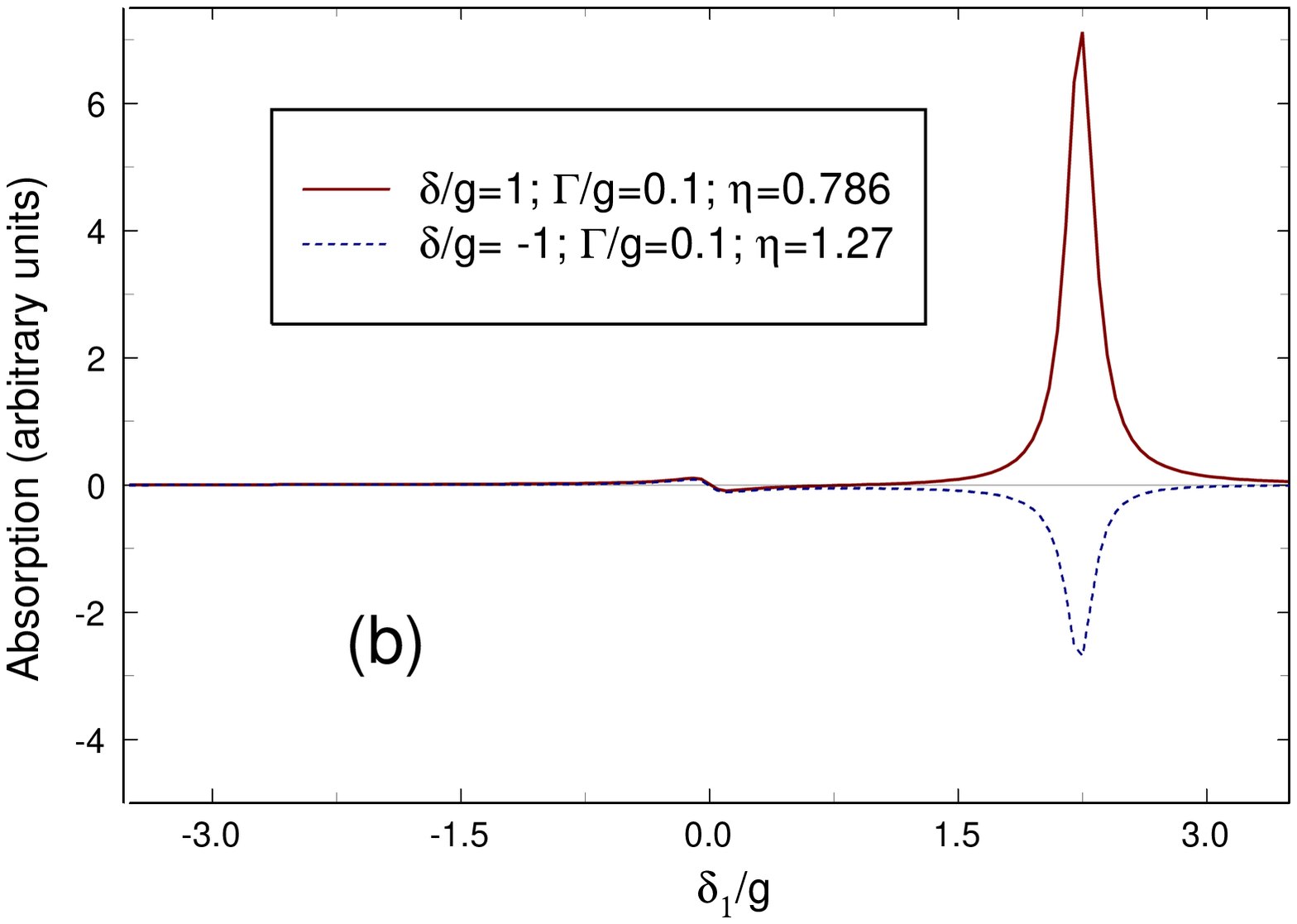}
\end{center}
\end{minipage}
\end{figure}
\begin{figure}
\begin{minipage}{.95\linewidth}
\begin{center}
\epsfxsize=.95\linewidth \epsfysize=.77\linewidth \epsfbox{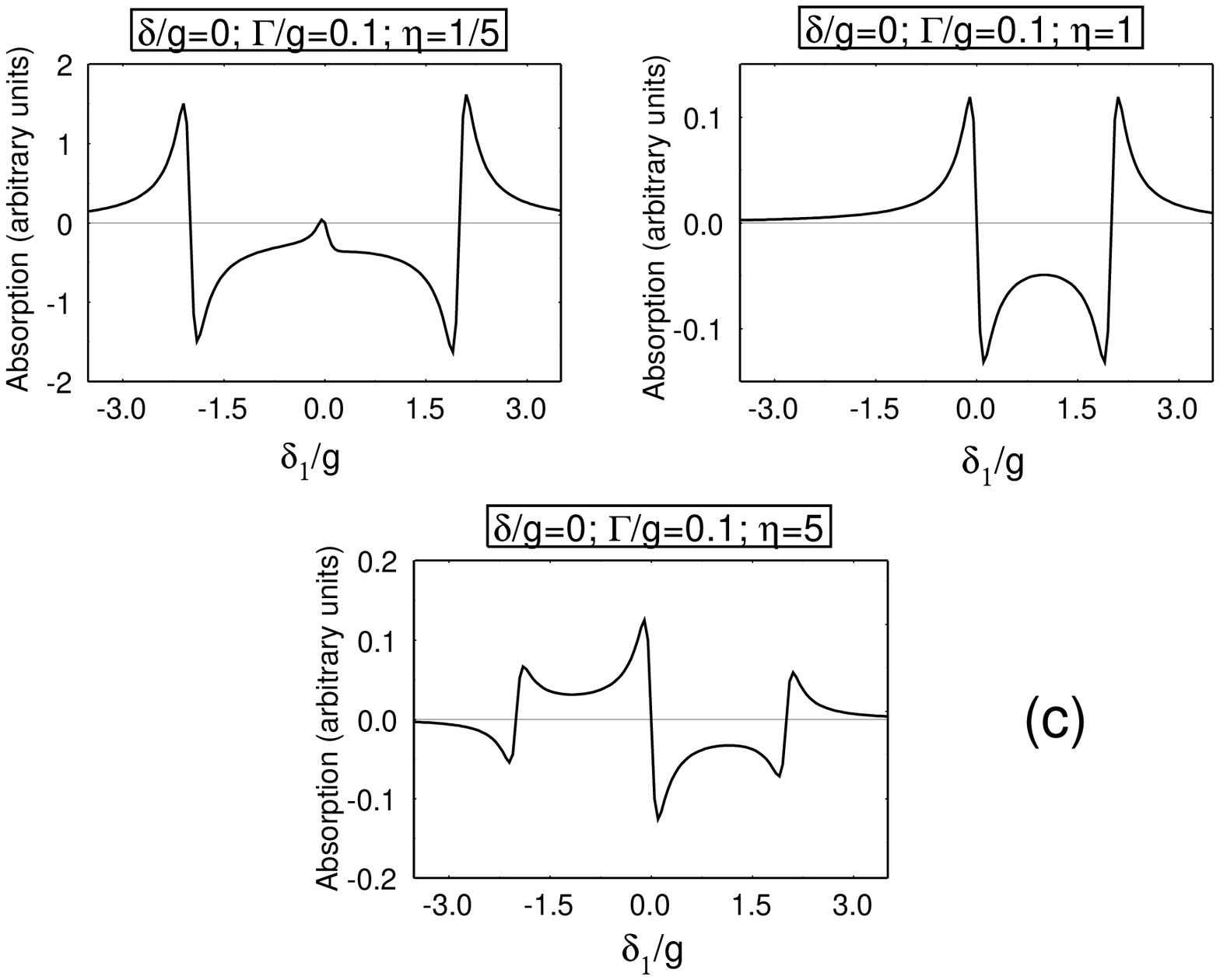}
\end{center}
\end{minipage}
\begin{minipage}{\linewidth} \caption{Probe field absorption in 
arbitrary units. Positive ordinate values
correspond to probe absorption and negative values to probe gain. 
\label{fig2}}
\end{minipage}
\end{figure}

The case of $\delta /g=0$ is shown in Fig. \ref{fig2}(c) for $\eta =1/5,1,$5, and $%
\Delta >0.$ If $\eta =1/5$, the spectrum is similar to that found for single
quantum transitions \cite{mol}. For $\eta =1,$ the spectral component at
negative $\delta _{1}$ is found to vanish. When $\eta \gtrsim 1$, there is a
dispersive-like structure centered at $\delta _{1}=0$ that is not found in
the pump-probe spectroscopy of single quantum transitions. Expressions for
the three components are given in Eqs. (A8) of Appendix A for $\left|
g\right| \gg \Gamma $, $\Gamma \ll \eta ^{2}$.

\section{Dressed atom approach}

The spectral features seen in Figs. \ref{fig2} (a),(b) can be explained using a
dressed atom approach. Semiclassical dressed states for two-quantum
transitions can be introduced via the transformation \cite{2} 
\begin{mathletters}
\label{14}
\begin{eqnarray}
\left( 
\begin{array}{c}
\left| A\right\rangle  \\ 
\left| B\right\rangle 
\end{array}
\right)  &=&{\bf T}\left( 
\begin{array}{c}
\left| 1\right\rangle  \\ 
\left| 2\right\rangle 
\end{array}
\right) ;  \label{14a} \\
{\bf T} &=&\left( 
\begin{array}{cc}
\cos \left( \theta \right)  & -\psi \sin \left( \theta \right)  \\ 
\psi \sin \left( \theta \right)  & \cos \left( \theta \right) 
\end{array}
\right) ,  \label{14b}
\end{eqnarray}
where 
\end{mathletters}
\begin{equation}
\omega _{BA}=\sqrt{\delta ^{2}+4g^{2}}  \label{15}
\end{equation}
is the frequency separation of the dressed states, 
\begin{equation}
\cos \left( \theta \right) =\left[ \frac{1}{2}\left( 1+\frac{\delta }{\omega
_{BA}}\right) \right] ^{1/2},  \label{16}
\end{equation}
and 
\begin{equation}
\psi =\left| \Delta \right| /\Delta .  \label{17}
\end{equation}
The angle $\theta $ is restricted such that $0\leq \theta \leq \pi /4$ for $%
\delta >0$ and $\pi /4\leq \theta \leq \pi /2$ for $\delta <0$. For $\theta
\sim 0$ ($\delta >0$, $\left| g/\delta \right| \ll 1$)$,$ $\left|
A\right\rangle \sim \left| 1\right\rangle $, while for $\theta \sim \pi /2$ $%
(\delta <0,$ $\left| g/\delta \right| \ll 1),$ $\left| B\right\rangle \sim
\left| 1\right\rangle $. In the secular approximation, 
\begin{equation}
\Gamma \ll \omega _{BA},  \label{18}
\end{equation}
it follows from Eqs. (\ref{9}) and (\ref{14}) that, to zeroth order in the
probe field, the diagonal dressed state density matrix elements are given by 
\begin{mathletters}
\label{19}
\begin{eqnarray}
&&\left. \rho _{AA}^{(0)}=(\Lambda _{1}/\Gamma )\cos ^{2}(\theta )+(\Lambda
_{2}/\Gamma )\sin ^{2}(\theta )\equiv \Lambda _{A}/\Gamma ;\right. 
\label{19a} \\
&&\left. \rho _{BB}^{(0)}=(\Lambda _{2}/\Gamma )\cos ^{2}(\theta )+(\Lambda
_{1}/\Gamma )\sin ^{2}(\theta )\equiv \Lambda _{B}/\Gamma ;\right. 
\label{19b} \\
&&\left. \rho _{AA}^{(0)}-\rho _{BB}^{(0)}=\left( \Lambda _{A}-\Lambda
_{B}\right) /\Gamma \right.   \nonumber \\
&&\left. =\left[ (\Lambda _{1}-\Lambda _{2})/\Gamma \right] \cos (2\theta
);.\right.   \label{19c}
\end{eqnarray}
Note that $\left( \rho _{AA}^{(0)}-\rho _{BB}^{(0)}\right) $ has the same
sign as $(\Lambda _{1}-\Lambda _{2})$ if $\delta >0$ and the opposite sign
if $\delta <0.$
\begin{figure}
\begin{minipage}{.95\linewidth}
\begin{center}
\epsfxsize=.95\linewidth \epsfysize=.66\linewidth \epsfbox{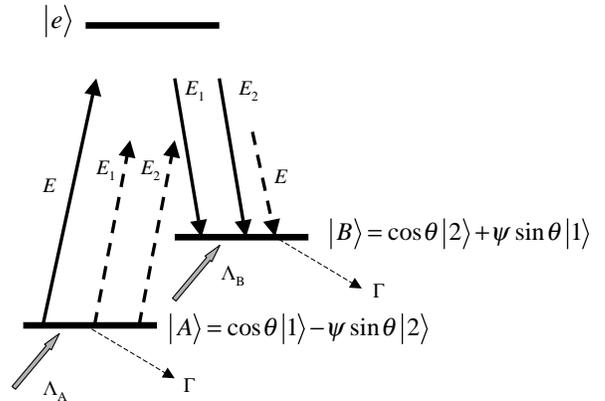}
\end{center}
\end{minipage}
\begin{minipage}{\linewidth} \caption{Dressed-state energy level 
diagram. In the interaction
representation adopted in the text, the frequency of field $E_{2}$ must be
set equal to $\Omega _{1}$ in calculating resonance conditions. For $\left(
\Lambda _{A}-\Lambda _{B}\right) >0$, solid arrows correspond to probe
absorption centered at $\protect\delta _{1}=\protect\omega _{BA}$ and dashed
arrows correspond to probe gain centered at $\protect\delta _{1}=-\protect%
\omega _{BA}$. 
\label{fig3}}
\end{minipage}
\end{figure}

It is now possible to use the energy level diagram (Fig. \ref{fig3}) to read directly
the probe absorption spectrum. The probe field is absorbed (or amplified)
via two quantum transitions between states $\left| A\right\rangle $ and $%
\left| B\right\rangle $. The two quantum transitions involve one photon from
the probe field and one photon from {\em either} field $E_{1}$ or $E_{2}$,
since all of these fields couple states $\left| A\right\rangle $ and $\left|
B\right\rangle $ to state $\left| e\right\rangle $. It is important to
remember that the effective field frequency of field $E_{2}$ is equal to $%
\Omega _{1}$ in this interaction representation. Fields $E_{1}$ and $E$
couple state $\left| e\right\rangle $ to the components of states $\left|
A\right\rangle $ and $\left| B\right\rangle $ involving state $\left|
1\right\rangle $, while field $E_{2}$ couples state $\left| e\right\rangle $
to the components of states $\left| A\right\rangle $ and $\left|
B\right\rangle $ involving state $\left| 2\right\rangle $. For example the
matrix element for the two-quantum process from state $\left| A\right\rangle 
$ to $\left| B\right\rangle $ involving absorption of a probe photon and
emission of a field $E_{2}$ photon is 
\end{mathletters}
\[
\frac{-i\chi }{-i\Delta }\cos (\theta )\frac{(-i\chi _{2})}{\Gamma -i(\delta
_{1}-\omega _{BA})}\cos (\theta ), 
\]
while that for absorption of a probe photon and emission of a field $E_{1}$
photon is 
\[
\frac{-i\chi }{-i\Delta }\cos (\theta )\frac{(-i\chi _{1})}{\Gamma -i(\delta
_{1}-\omega _{BA})}\psi \sin (\theta ). 
\]
These two processes add coherently, such that probe absorption via
transitions from state $\left| A\right\rangle $ to $\left| B\right\rangle $
is proportional to the sum of these two matrix elements squared, multiplied
by the population difference $\left( \rho _{AA}^{(0)}-\rho
_{BB}^{(0)}\right) .$ In other words, the probe absorption at $\delta
_{1}=\omega _{BA}$ is proportional to a quantity $C_{+}$ given by

\end{multicols}

\begin{equation}
C_{+}=\left( g/\Delta \Gamma \right) \left[ (\Lambda _{1}-\Lambda
_{2})/\Gamma \right] \cos \left( 2\theta \right) \left( \psi \eta \sin
\left( \theta \right) \cos (\theta )+\frac{1}{\eta }\cos ^{2}\left( \theta
\right) \right) ^{2}.  \label{20}
\end{equation}
Similarly, probe gain via transitions from state $\left| A\right\rangle $ to 
$\left| B\right\rangle $ at $\delta _{1}=-\omega _{BA}$ is proportional to 
\begin{equation}
C_{-}=\left( g/\Delta \Gamma \right) \left[ (\Lambda _{1}-\Lambda
_{2})/\Gamma \right] \cos \left( 2\theta \right) )\left( \psi \eta \sin
\left( \theta \right) \cos \left( \theta \right) -\frac{1}{\eta }\sin
^{2}\left( \theta \right) \right) ^{2}.  \label{21}
\end{equation}

\begin{multicols}{2}

A formal derivation of these results is given in Appendix B.

\begin{figure}
\begin{minipage}{.95\linewidth}
\begin{center}
\epsfxsize=.95\linewidth \epsfysize=.91\linewidth \epsfbox{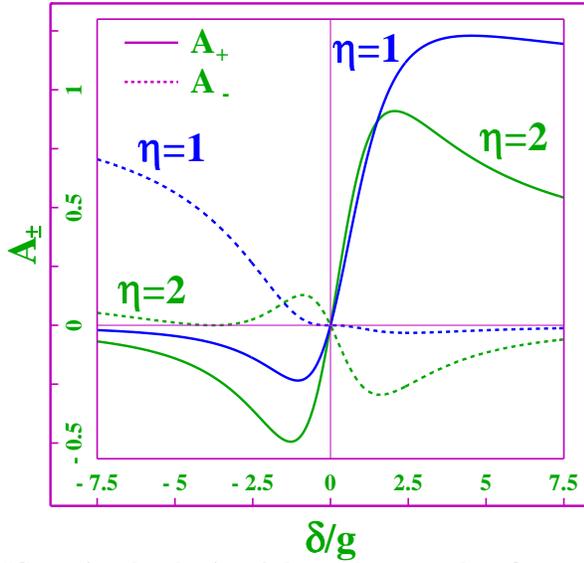}
\end{center}
\end{minipage}
\begin{minipage}{\linewidth} \caption{Amplitude $A_{+}$ of the peak
 centered at $\protect\delta _{1}=%
\protect\omega _{BA}$ and amplitude $A_{-}$ of the peak centered at $\protect%
\delta _{1}=-\protect\omega _{BA}$, for $\Delta >0.$ Positive values of $%
A_{\pm }$ correspond to absorption and negative values to gain. 
\label{fig4}}
\end{minipage}
\end{figure}

For the sake of definiteness, let us take $(\Lambda _{1}-\Lambda _{2})>0;$
then $C_{+}$ corresponds to absorption for $\delta >0$ and to gain for $%
\delta <0$, while $C_{-}$ corresponds to gain for $\delta >0$ and to
absorption for $\delta <0$. Note that the component centered at $\delta
_{1}=-\omega _{BA}$ vanishes if $\Delta >0$ and $\tan (\theta )=\eta ^{2}$,
while that at $\delta _{1}=\omega _{BA}$ vanishes if $\Delta <0$ and $\tan
(\theta )=\eta ^{-2}$. The values of $A_{\pm }=\pm C_{\pm }\left[ \Gamma
^{2}\Delta /\left| g\right| (\Lambda _{1}-\Lambda _{2})\right] $ are plotted
in Fig. \ref{fig4} as a function of $\delta /g$ for $\Delta >0$ and $\eta =1,2$. For $%
\Delta <0$, one can use the relationship $A_{\pm }(-\Delta ,-\delta )=A_{\mp
}(\Delta ,\delta )$.

The probe absorption vanishes in the secular approximation (\ref{18}) when $%
\delta =0,$ since, in this case, $\theta =\frac{\pi }{4}$ and the
populations of the dressed states\ are equal. The lowest order dressed atom
approach is not useful in this limit. Typical spectra are shown in Fig. \ref{fig2}(c) 
and were discussed in Sec. III$.$

\section{Conclusion}

The probe absorption spectrum has been calculated for two-quantum
transitions between levels that are simultaneously driven by a two-quantum
pump field of arbitrary intensity. In addition to features found in
conventional pump-probe spectroscopy of single quantum transitions, new
features have been found that can be identified with interference phenomena.
Both Doppler and recoil effects were neglected in out treatment. For nearly
copropagating fields, effects arising from these processes are negligible.
Doppler shifts can be accounted for by the replacements $\delta
_{1}\rightarrow \delta _{1}+({\bf k}_{1}-{\bf k})\cdot {\bf v}$, $\delta
_{1}-\tilde{\delta}\rightarrow \delta _{1}-\tilde{\delta}+({\bf k}_{2}-{\bf k%
})\cdot {\bf v,}$ and $\delta _{1}+\tilde{\delta}\rightarrow \delta _{1}+%
\tilde{\delta}+(2{\bf k}_{1}-{\bf k}_{2}-{\bf k})\cdot {\bf v}$ in the
equations in the Appendix.

\begin{figure}
\begin{minipage}{.45\linewidth}
\begin{center}
\epsfxsize=.95\linewidth \epsfysize=.66\linewidth \epsfbox{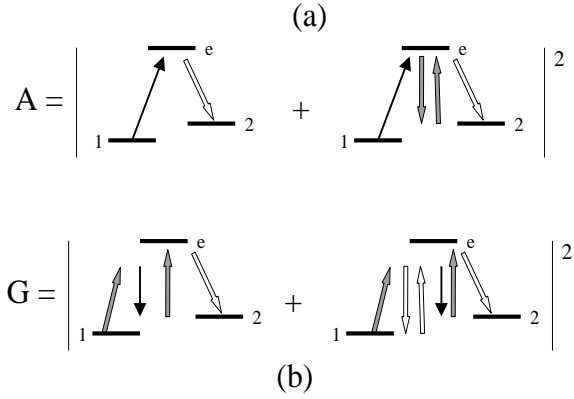}
\end{center}
\end{minipage}
\begin{minipage}{.45\linewidth} \caption{Schematic representation of 
the 1$\rightarrow 2$ transition
probability leading to probe absorption or probe gain in lowest order
perturbation theory in the bare basis. The thin arrow represents the probe
field, the broad filled arrows field $E_{1}$, and the broad open arrows
field $E_{2}$. (a) absorption, (b) gain. Terms involving the sequential
absorption and emission of the {\em same} field have been neglected, since
such terms result only in Stark shifts of levels $1$ and $2$. The diagrams
are drawn for $\tilde{\protect\delta}>0$; if $\tilde{\protect\delta}<0$, the
roles of absorption and gain would be interchanged.
\label{fig5}}
\end{minipage}
\end{figure}

The dependence of the interference effect of the signs of $\Delta $ and $%
\tilde{\delta}$ can be understood in the bare atom picture in a perturbative
limit. A schematic representation of the probability amplitude leading to
probe absorption at $\delta _{1}=\tilde{\delta}$ is shown in Fig. \ref{fig5}(a). Each
arrow represents an interaction with one of the fields. The two
contributions to the final state amplitude add coherently. Putting in the
appropriate energy denominators, one finds that the absorption varies as

\end{multicols}

\begin{equation}
A=\left| \frac{i^{2}\chi \chi _{2}^{\ast }}{\left( \gamma _{e}/2-i\Delta
\right) \left[ \Gamma -i\left( \Delta -\Delta _{2}\right) \right] }+\frac{%
i^{4}\chi \chi _{2}^{\ast }\left| \chi _{1}\right| ^{2}}{\left( \gamma
_{e}/2-i\Delta \right) \left[ \Gamma -i\left( \Delta -\Delta _{1}\right) %
\right] \left( \gamma _{e}/2-i\Delta \right) \left[ \Gamma -i\left( \Delta
-\Delta _{2}\right) \right] }\right| ^{2}.  \label{22}
\end{equation}

\begin{multicols}{2}

For $\left| \tilde{\delta}\right| \gg \Gamma ,$ and $\left| \Delta \right|
\gg \gamma _{e}$, this equation reduces to 
\begin{equation}
A=\left| \frac{\chi \chi _{2}^{\ast }}{\Delta }\right| ^{2}\frac{1}{\Gamma
^{2}+\left( \delta _{1}-\tilde{\delta}\right) ^{2}}\left| 1+\frac{\left|
g\right| \eta ^{2}\psi }{\tilde{\delta}}\right| ,  \label{23}
\end{equation}
which shows the dependence on the signs of $\Delta $ ($\psi =\left| \Delta
\right| /\Delta )$ and $\tilde{\delta}.$ A similar calculation for the
emission component represented schematically in Fig. \ref{fig5}(b) leads to 
\begin{equation}
G=\left| \frac{\chi ^{\ast }\chi _{1}^{2}\chi _{2}^{\ast }}{\Delta ^{2}%
\tilde{\delta}}\right| ^{2}\frac{1}{\Gamma ^{2}+\left( \delta _{1}+\tilde{%
\delta}\right) ^{2}}\left| 1-\frac{\left| g\right| \eta ^{-2}\psi }{\tilde{%
\delta}}\right| .  \label{24}
\end{equation}

New effects will arise if the fields are not copropagating and the active
medium is a subrecoil cooled atomic 

\end{multicols}

\begin{multicols}{2}
vapor, a highly collimated atomic beam,
or a BEC. As for single quantum transitions \cite{recoil}, each component of
the spectrum undergoes recoil splitting. Since the center-of-mass momentum
states differ for two-quantum processes involving fields $E_{1}$ and $E$
from those involving fields $E_{2}$ and $E$, one might expect the spectrum
consists of eight absorption and eight emission components rather then the
four absorption and four emission components found for single quantum
transitions; however this does not appear to be the case. Instead, each
component results from a coherent superposition of two quantum processes
involving fields $\left( E_{1},E\right) $ and $\left( E_{2},E\right) $.

\section{Acknowledgments}

This work is supported by the U. S. Army Research Office under Grant No.
DAAG55-97-0113 and by the National Science Foundation under Grant No.
PHY-9800981. We are grateful to the Prof. G. Raithel for fruitful
discussions.

\end{multicols}

$\_\_\_\_\_\_\_\_\_\_\_\_\_\_\_\_\_\_\_\_\_\_\_\_\_\_\_\_\_\_\_\_\_\_\_\_\_%
\_\_\_\_\_\_\_\_\_\_\_\_\_\_\_\_\_\_\_\_\_\_\_\_\_\_\_\_\_\_\_\_\_\_\_\_\_%
\_\_\_\_\_\_\_\_\_\_\_\_\_\_\_\_\_\_\_\_\_\_\_\_\_\_\_\_\_\_\_%
\_\_\_\_\_\_\_\_\_\_\_\_\_\_\_\_\_\_\_\_\_\_\_\_\_\_\_\_\_\_\_$

\appendix

\begin{multicols} {2}

\section{Bare State Calculations}

Substituting Eqs. (\ref{12}) into Eqs. (\ref{9}), one finds to zeroth order
in the probe field that 
\begin{mathletters}
\label{A1}
\begin{eqnarray}
w^{(0)} &=&\rho _{22}^{(0)}-\rho _{11}^{(0)}=\frac{(\Lambda _{2}-\Lambda
_{1})}{\Gamma }\frac{\Gamma ^{2}+\delta ^{2}}{\Gamma ^{2}+\delta
^{2}+4\left| g\right| ^{2}};  \label{A1a} \\
\rho _{11}^{(0)} &=&\frac{1}{2}\left[ \frac{(\Lambda _{2}+\Lambda _{1})}{%
\Gamma }-w^{(0)}\right] ;  \label{A1b} \\
\rho _{12}^{(0)} &=&\frac{-ig}{\Gamma -i\delta }w^{(0)}=\left( \rho
_{21}^{(0)}\right) ^{\ast },  \label{A1c}
\end{eqnarray}
and that, to first order in the probe field, $w^{+}=\rho _{22}^{+}-\rho
_{11}^{+}$, $\rho _{12}^{+}$, $\rho _{21}^{+}$, and $m^{+}=\rho
_{22}^{+}+\rho _{11}^{+}$ satisfy 
\end{mathletters}
\begin{mathletters}
\label{A2}
\begin{eqnarray}
&&\left. m^{+}=0;\right.   \label{A2a} \\
&&\left. \left( \Gamma +i\delta _{1}\right) w^{+}-2ig^{\ast }\rho
_{21}^{+}+2ig\rho _{12}^{+}=2ig^{\prime }\rho _{21}^{(0)};\right. 
\label{A2b} \\
&&\left. \left[ \Gamma +i\left( \delta _{1}-\delta \right) \right] \rho
_{12}^{+}+ig^{\ast }w^{+}=-ig^{\prime \ast }w^{(0)}-iS\rho
_{12}^{(0)};\right.   \label{A2c} \\
&&\left. \left[ \Gamma +i\left( \delta _{1}+\delta \right) \right] \rho
_{21}^{+}-igw^{+}=iS\rho _{21}^{(0)}.\right.   \label{A2d}
\end{eqnarray}
Equation (\ref{11}) can be rewritten as 
\end{mathletters}
\begin{equation}
\rho _{1e}^{\prime }\approx (\chi ^{\ast }/\Delta )\left[ \rho _{11}^{(0)}-%
\frac{1}{2}y_{0}+y_{12}\right] ,  \label{A3}
\end{equation}
where 
\begin{mathletters}
\label{A4}
\begin{eqnarray}
y_{0} &=&\left( \chi _{1}/\chi \right) ^{\ast }w^{+};  \label{A4a} \\
y_{12} &=&\left( \chi _{2}/\chi \right) ^{\ast }\rho _{12}^{+};  \label{A4b}
\\
y_{21} &=&\left( \frac{\chi _{2}\chi _{_{1}}^{\ast }}{\chi ^{\ast }\chi _{1}}%
\right) \rho _{21}^{+}.  \label{A4c}
\end{eqnarray}
In Eqs. (\ref{A1}-\ref{A4}) we have allowed the Rabi frequencies to be
complex.

The quantities $y_{0}$, $y_{12}$, and $y_{21}$ satisfy the coupled
equations: 
\end{mathletters}
\begin{mathletters}
\label{A5}
\begin{eqnarray}
\left[ \Gamma +i\left( \delta _{1}-\delta \right) \right] y_{12}+i\left|
g\right| \psi \eta ^{-2}y_{0} &=&a;  \label{A5a} \\
\left( \Gamma +i\delta _{1}\right) y_{0}-2i\left| g\right| \psi \eta
^{2}y_{21}+2i\left| g\right| \psi \eta ^{2}y_{12} &=&b;  \label{A5b} \\
\left[ \Gamma +i\left( \delta _{1}+\delta \right) \right] y_{21}-i\left|
g\right| \psi \eta ^{-2}y_{0} &=&c,  \label{A5c}
\end{eqnarray}
where 
\end{mathletters}
\begin{mathletters}
\label{A6}
\begin{eqnarray}
b &=&\frac{-2\left| g\right| ^{2}}{\Gamma +i\delta }w^{(0)}=2c  \label{A6a}
\\
a &=&-i\left| g\right| \psi \eta ^{-2}w^{(0)}-\frac{\left| g\right| ^{2}}{%
\Gamma -i\delta }w^{(0)}.  \label{A6b}
\end{eqnarray}
and $\psi =(\left| \Delta \right| /\Delta )$. Note that the equations do not
depend on the phase of the various Rabi frequencies, but {\em do} depend on
the sign of $\Delta $. Explicit solutions for $y_{0}$ and $y_{12}$ are: 
\end{mathletters}

\end{multicols}

\begin{mathletters}
\label{A7}
\begin{eqnarray}
y_{0} &=&-i\frac{2a\left| g\right| \psi \eta ^{2}\left( \delta +\delta
_{1}-i\Gamma \right) +b\left[ \delta ^{2}-\left( \delta _{1}-i\Gamma \right)
^{2}\right] +2c\left| g\right| \psi \eta ^{2}\left( \delta -\delta
_{1}+i\Gamma \right) }{\left( \delta _{1}-i\Gamma \right) \left( \delta
^{2}-\delta _{1}^{2}+2i\Gamma \delta _{1}+\Gamma ^{2}+4\left| g\right|
^{2}\right) };  \label{A7a} \\
y_{12} &=&i\frac{a\left[ \delta _{1}^{2}+\delta \left( \delta _{1}-i\Gamma
\right) -2i\Gamma \delta _{1}-\Gamma ^{2}-2\left| g\right| ^{2}\right]
-b\eta ^{-2}\left| g\right| \psi \left( \delta +\delta _{1}-i\Gamma \right)
-2c\left| g\right| ^{2}}{\left( \delta _{1}-i\Gamma \right) \left( \delta
^{2}-\delta _{1}^{2}+2i\Gamma \delta _{1}+\Gamma ^{2}+4\left| g\right|
^{2}\right) }.  \label{A7b}
\end{eqnarray}
\end{mathletters}

\begin{multicols}{2}

The line shape is totally non-secular when $\delta =0.$ In the limit that $%
\Delta >0$, $\left| g\right| \gg \Gamma $, and $\Gamma \ll \eta ^{2},$ one
finds that the absorption coefficient $\alpha $ for $\delta _{1}\approx 0$
is 
\begin{equation}
\alpha \sim -\frac{1}{4}\left( \frac{kNd_{1e}^{2}}{2\hbar \epsilon
_{0}\Delta }\right) \frac{\Lambda _{1}-\Lambda _{2}}{\Gamma }\frac{\delta
_{1}\Gamma }{\left( \delta _{1}^{2}+\Gamma ^{2}\right) },  \eqnum{A8a}
\label{A8a}
\end{equation}
that for $\delta _{1}\approx 2\left| g\right| $ is 
\begin{equation}
\alpha \sim \frac{1}{8}\left( \frac{kNd_{1e}^{2}}{2\hbar \epsilon _{0}\Delta 
}\right) \frac{\Lambda _{1}-\Lambda _{2}}{\Gamma }\frac{\left( \delta
_{1}-2\left| g\right| \right) \Gamma }{\left[ \left( \delta _{1}-2\left|
g\right| \right) ^{2}+\Gamma ^{2}\right] }(1+\eta ^{-2}),  \eqnum{A8b}
\label{A8b}
\end{equation}
and that for $\delta _{1}\approx -2\left| g\right| $ is 
\begin{equation}
\alpha \sim \frac{1}{8}\left( \frac{kNd_{1e}^{2}}{2\hbar \epsilon _{0}\Delta 
}\right) \frac{\Lambda _{1}-\Lambda _{2}}{\Gamma }\frac{\left( \delta
_{1}+2\left| g\right| \right) \Gamma }{\left[ \left( \delta _{1}+2\left|
g\right| \right) ^{2}+\Gamma ^{2}\right] }(1-\eta ^{-2}).  \eqnum{A8c}
\label{A8c}
\end{equation}
Note that the component at $\delta _{1}=-2\left| g\right| $ vanishes if $%
\eta =1$. For $\Delta <0$, one can use Eq. (\ref{reflect}).

\section{Dressed-State Calculations}

Equation (\ref{7}) can be written in the form 
\begin{equation}
i\hbar {\bf \dot{b}}=\left( {\bf V+V}_{p}\right) {\bf b},  \label{B1}
\end{equation}
where 
\begin{equation}
{\bf V}=\hbar \left( 
\begin{array}{cc}
-\delta /2 & g^{\ast } \\ 
g & \delta /2
\end{array}
\right) ,  \label{B2}
\end{equation}
\begin{equation}
{\bf V}_{p}=\hbar \left( 
\begin{array}{cc}
Se^{i\delta _{1}t}+S^{\ast }e^{-i\delta _{1}t} & g^{\prime \ast }e^{i\delta
_{1}t} \\ 
g^{\prime }e^{-i\delta _{1}t} & 0
\end{array}
\right) ,  \label{B3}
\end{equation}
\begin{equation}
g=\frac{\chi _{1}\chi _{2}^{\ast }}{\Delta };\text{ \ }g^{\prime }=\frac{%
\chi \chi _{2}^{\ast }}{\Delta };\text{ \ }S=\frac{\chi ^{\ast }\chi _{1}}{%
\Delta },  \label{B5}
\end{equation}
and we have allowed for complex Rabi frequencies, 
\begin{equation}
\chi _{1}=\left| \chi _{1}\right| e^{i\phi _{1}},\chi _{2}=\left| \chi
_{2}\right| e^{i\phi _{2}},\chi =\left| \chi \right| e^{i\phi }.  \label{B6}
\end{equation}

If one introduces semi-classical dressed states via the transformation 
\begin{equation}
{\bf b}_{d}={\bf T}_{c}{\bf b},  \label{B7}
\end{equation}
where 
\begin{equation}
{\bf b}_{d}=\left( 
\begin{array}{c}
A \\ 
B
\end{array}
\right) ,  \label{B8}
\end{equation}
\begin{equation}
{\bf T}_{c}=\left( 
\begin{array}{cc}
\cos \left( \theta \right) e^{i\phi _{d}/2} & -e^{-i\phi _{d}/2}\sin \left(
\theta \right) \\ 
e^{i\phi _{d}/2}\sin \left( \theta \right) & e^{-i\phi _{d}/2}\cos \left(
\theta \right)
\end{array}
\right) ,  \label{B9}
\end{equation}
and 
\begin{equation}
\phi _{d}=\phi _{1}-\phi _{2}+\frac{\pi }{2}(1-\psi )  \label{B10}
\end{equation}
(recall that $\psi =\left| \Delta \right| /\Delta $), then the dressed-state
Hamiltonian is given by 
\begin{equation}
{\bf V}_{d}=\hbar \left( 
\begin{array}{cc}
-\omega _{BA}/2 & 0 \\ 
0 & \omega _{BA}/2
\end{array}
\right) +{\bf T}_{c}{\bf V}_{p}{\bf T}_{c}^{{\bf \dagger }}.  \label{B11}
\end{equation}
The dressed state density matrix, 
\begin{equation}
{\bf \rho }_{d}=\left( 
\begin{array}{cc}
\rho _{AA} & \rho _{AB} \\ 
\rho _{BA} & \rho _{BB}
\end{array}
\right)  \label{B12}
\end{equation}
evolves as 
\begin{equation}
\left( \frac{d}{dt}+\Gamma \right) {\bf \rho }_{d}\approx \frac{1}{i\hbar }%
\left[ {\bf V}_{d},{\bf \rho }_{d}\right] +\left( 
\begin{array}{cc}
\Lambda _{A} & 0 \\ 
0 & \Lambda _{B}
\end{array}
\right) ,  \label{B13}
\end{equation}
Off-diagonal terms have been neglected in the matrix representing the
incoherent pumping, since they give rise to terms of order $\Gamma /\omega
_{BA}\ll 1$ (secular approximation).

The dressed state density matrix is expanded as 
\begin{equation}
{\bf \rho }_{d}={\bf \rho }_{d}^{(0)}+{\bf \rho }_{d}^{+}e^{i\delta _{1}t}+%
{\bf \rho }_{d}^{-}e^{-i\delta _{1}t},  \label{B14}
\end{equation}
and it is found from Eqs. (\ref{B1})-(\ref{B3}), (\ref{B7})-(\ref{B14}) that 
${\bf \rho }_{d}^{+}$ obeys the equation of motion

\end{multicols}

\begin{equation}
\left( \frac{d}{dt}+\Gamma \right) {\bf \rho }_{d}^{+}=i\left( 
\begin{array}{cc}
0 & \left( \omega _{BA}-\delta _{1}\right) \rho _{AB}^{+} \\ 
-\left( \omega _{BA}+\delta _{1}\right) \rho _{BA}^{+} & 0
\end{array}
\right) +\frac{1}{i\hbar }\left[ {\bf V}_{pd},{\bf \rho }_{d}^{(0)}\right] ,
\label{B15}
\end{equation}
where 
\begin{equation}
{\bf V}_{pd}=\hbar \left( 
\begin{array}{cc}
\cos \left( \theta \right) \left[ S\cos \left( \theta \right) -g^{\prime
\ast }e^{i\phi _{d}}\sin \left( \theta \right) \right] ; & \cos \left(
\theta \right) \left[ S\sin \left( \theta \right) +g^{\prime \ast }\cos
\left( \theta \right) e^{i\phi _{d}}\right] \\ 
\sin \left( \theta \right) \left[ -g^{\prime \ast }\sin \left( \theta
\right) e^{i\phi _{d}}+S\cos \left( \theta \right) \right] ; & \sin \left(
\theta \right) \left[ S\sin \left( \theta \right) +g^{\prime \ast }\cos
\left( \theta \right) e^{i\phi _{d}}\right]
\end{array}
\right)  \label{B16}
\end{equation}
In the secular approximation, the steady state solution of Eq. (\ref{B15})
is 
\begin{equation}
{\bf \rho }_{d}^{+}=\left( 
\begin{array}{cc}
0 & \rho _{AB}^{+} \\ 
\rho _{BA}^{+} & 0
\end{array}
\right) ,  \label{B17}
\end{equation}
where 
\begin{mathletters}
\label{B18}
\begin{eqnarray}
\rho _{AB}^{+} &=&i\cos \left( \theta \right) \left[ S\sin \left( \theta
\right) +g^{\prime \ast }\cos \left( \theta \right) e^{i\phi _{d}}\right]
\left( \rho _{AA}^{(0)}-\rho _{BB}^{(0)}\right) /\left( \Gamma +i\left(
\delta _{1}-\omega _{BA}\right) \right) ,  \label{B18a} \\
\rho _{BA}^{+} &=&-i\sin \left( \theta \right) \left[ -g^{\prime \ast }\sin
\left( \theta \right) e^{i\phi _{d}}+S\cos \left( \theta \right) \right]
\left( \rho _{AA}^{(0)}-\rho _{BB}^{(0)}\right) /\left( \Gamma +i\left(
\delta _{1}+\omega _{BA}\right) \right) .  \label{B18b}
\end{eqnarray}

The coherence $\rho _{1e}^{\prime }$ needed in Eq. (\ref{10}) for the
absorption coefficient and index change is given by 
\end{mathletters}
\begin{equation}
\rho _{1e}^{\prime }\approx (1/\Delta )\left[ \chi ^{\ast }\rho
_{11}^{(0)}+\chi _{1}^{\ast }\rho _{11}^{+}+\chi _{2}^{\ast }\rho _{12}^{+}%
\right] .  \label{B19}
\end{equation}
The first term can be evaluated using Eq. (\ref{A1b}) for $\rho _{11}^{(0)}$%
; it contributes to the index change, but not the absorption. For the
remaining terms, one rewrites $\rho _{11}^{+}$ and $\rho _{12}^{+}$ in the
dressed basis using Eqs. (\ref{B7}),(\ref{B9}),(\ref{B12}), and uses Eq. (%
\ref{B6}) to extract all the phase factors to arrive at 
\begin{equation}
\rho _{1e}^{\prime }\approx (\chi ^{\ast }/\left| \Delta \right| )\left[
\psi \rho _{11}^{(0)}+f_{+}+f_{-}\right]  \label{B20}
\end{equation}
where 
\begin{mathletters}
\label{B21}
\begin{eqnarray}
f_{+} &=&\frac{i\left| g\right| }{\left[ \Gamma +i\left( \delta _{1}-\omega
_{BA}\right) \right] }\cos \left( 2\theta \right) \frac{\left( \Lambda
_{1}-\Lambda _{2}\right) }{\Gamma }\cos ^{2}(\theta )\left( \psi \eta \sin
\left( \theta \right) +\frac{1}{\eta }\cos \left( \theta \right) \right)
^{2},  \label{B21a} \\
f_{-} &=&-\frac{i\left| g\right| }{\left[ \Gamma +i\left( \delta _{1}+\omega
_{BA}\right) \right] }\cos \left( 2\theta \right) \frac{\left( \Lambda
_{1}-\Lambda _{2}\right) }{\Gamma }\sin ^{2}(\theta )\left( \psi \eta \cos
\left( \theta \right) -\frac{1}{\eta }\sin \left( \theta \right) \right)
^{2}.  \label{B21b}
\end{eqnarray}

Note that the approach and results of Sec. III are unchanged if one uses
complex dressed states defined by 
\end{mathletters}
\begin{equation}
\left( 
\begin{array}{c}
\left| A\right\rangle \\ 
\left| B\right\rangle
\end{array}
\right) ={\bf T}_{c}^{\ast }\left( 
\begin{array}{c}
\left| 1\right\rangle \\ 
\left| 2\right\rangle
\end{array}
\right) .  \label{B22}
\end{equation}

\begin{multicols}{2}

\end{multicols}

\end{document}